\documentclass[aps,prd,eqsecnum]{revtex4}

\usepackage{graphicx}
\usepackage{lscape}
\usepackage{indentfirst}
\usepackage{latexsym}
\usepackage{multirow}
\usepackage{tabls}
\usepackage{epsfig}
\usepackage{color}
\usepackage{amssymb}
\usepackage{subfigure}

\usepackage{amsfonts}
\usepackage{amsmath}
\usepackage{bm}
\usepackage{mathrsfs}

\def\barh{ {\bar h} }

\def\barx{ {\bar x} }

\def\cB{ {\cal B} }

\def\cE{ {\cal E} }
\def\cG{ {\cal G} }

\def\cT{ {\cal T} }

\def\br{ {\bf r} }
\def\bR{ {\bf R} }

\def\bV{ {\bf V} }
\def\bv{ {\bf v} }

\def\bz{ {\bf z} }
\def\bZ{ {\bf Z} }

\def\dz{ { \dot{z} } }

\def\bzero{ {\bm 0} }

\def\lsim{\mathrel{\rlap{\lower3pt\hbox{\hskip1pt$\sim$}}
    \raise1pt\hbox{$<$}}}                
\def\gsim{\mathrel{\rlap{\lower3pt\hbox{\hskip1pt$\sim$}}
    \raise1pt\hbox{$>$}}}         

\def\coordeq{ \, \mathrel{ \rlap{\hbox{\hskip-2.5pt$=$} }
    \raise4pt\hbox{$\cdot$}} \, }                

\begin{document}

\title{New analytical methods for gravitational radiation and reaction in binaries with arbitrary mass
ratio and relative velocity\footnote{Invited contribution to the
International Conference on Classical and Quantum Relativistic
Dynamics of Particles and Fields (IARD) held at the Aristotle
University, Thessaloniki, Greece, 22-26 June 2008. Proceedings to
appear in {\it Foundations of Physics}.}}
\author{Chad R. Galley$^{1,2}$\footnote{crgalley@umd.edu} and B. L. Hu$^{2}$\footnote{blhu@umd.edu}}

\affiliation{$^{1}$Center for Scientific Computation and Mathematical
Modeling, and \\
$^{2}$Maryland Center for Fundamental Physics, Department of Physics,
University of Maryland, College Park, MD, 20742}


\begin{abstract}
We present a new analytical framework for describing the dynamics of
a gravitational binary system with unequal masses moving with
arbitrary relative velocity, taking into account the backreaction
from both compact objects in the form of tidal deformation,
gravitational waves and self forces. Allowing all dynamical variables
to interact with each other in a self-consistent manner this
formalism ensures that all the dynamical quantities involved are
conserved on the background spacetime and obey the gauge invariance
under general coordinate transformations that preserve the background
geometry.
Because it is based on a generalized perturbation theory and the
important new emphasis is on the self-consistency of all the
dynamical variables involved we call it a gravitational perturbation
theory with self-consistent backreaction (GP-SCB).

As an illustration of how this formalism is implemented we construct
perturbatively a self-consistent set of equations of motion for an
inspiraling gravitational binary, which does not require extra
assumptions such as slow motion, weak-field or small mass ratio for
its formulation. This case should encompass the inspiral and possibly
the plunge and merger phases of binaries with otherwise general
parameters (e.g., mass ratio and relative velocity) though more
investigation is needed to substantiate it.

In the second part, we discuss how the mass ratio can be treated as a
perturbation parameter in the post-Newtonian effective field theory
(PN-EFT) approach,  thus extending the work of Goldberger and
Rothstein for equal mass binaries to variable mass ratios $\eta
\equiv m/M$ ($M \ge m$). As the mass ratio $\eta$ decreases from 1 to
near 0, the smaller mass can be accelerated to higher speeds thereby
requiring higher order post-Newtonian corrections to describe the
system with sufficient accuracy for the data analysis requirements of
current and future gravitational wave interferometer detectors. We
provide rough estimates for the higher post-Newtonian orders needed
{to determine the number of gravitational wave cycles, with a
specified precision, that fall into a detector's bandwidth. For
example, we find that the number of cycles may need to be calculated
through 4.5PN for $\sim 5\%$ precision (relative to 2.5PN
corrections) for a compact binary with $10 M_\odot$ and $100 M_\odot$
constituents.}

\end{abstract}

\maketitle

\section{Introduction}
\label{sec:intro}

Ground-based gravitational wave interferometer detectors,  such as
LIGO \cite{Abramovici:LIGO}, are currently operating at their
intended design sensitivity and are actively constraining the
amplitudes of gravitational waves that may be passing within their
detectable bandwidth. These observatories are expected to detect
gravitational wave signals from the inspiral, plunge, merger and
ringdown phases of binaries with astrophysical masses, ranging from
the mass of a neutron star to the mass of several hundred solar
masses. Space-based detectors, such as LISA \cite{LISA}, which are
currently in the planning and development stages are designed to
detect the gravitational waves from inspirals of extreme mass ratio
binaries and from binaries with nearly equal and large masses (e.g.,
with masses $\sim 10^6 M_\odot$), among other sources.

\subsection{Existing analytical formalisms for gravitational binaries}

Different descriptions of gravitational binary systems ultimately
depend on the parameters of the system. For example, the
post-Newtonian (PN) approximation (see below) is useful for binaries
with nearly equal masses ($\lsim 10 M_\odot$) during the inspiral
phase, with extreme mass ratios where the small compact object moves
in the weak-field region of the supermassive black hole, and with
supermassive black holes orbiting slowly in a weak-field regime. For
motion in a strong field, one often uses gravitational perturbation
(GP) theory on a fixed but curved background (see below) to describe
the perturbations of a small compact object moving in the background
field of a supermassive black hole. We review these formalisms in
somewhat more detail below.

\subsubsection{The post-Newtonian expansion}

The post-Newtonian approximation is based on the assumption that two
weakly gravitating objects orbit about each other at nonrelativistic
speeds on a flat background. The strict weak field requirement  can
be lifted by constructing an annulus about each compact object where
the PN-expanded metric is matched onto the near-field perturbed
metric of the strongly gravitating compact object
\cite{Damour:300yrs, Thorne:RevModPhys52}.

Iteratively solving for the metric  perturbations and the positions
of the masses yields approximate expressions in powers of the
relative velocity for the phase of the emitted gravitational
radiation, the energy and angular momentum they carry, the location
of the innermost stable circular orbit, etc. To date, the equations
of motion for the compact objects have been computed to order $v^7$
beyond the Newtonian ones, also denoted as 3.5PN, for non-spinning
compact objects. For good reviews of the PN expansion see
\cite{Blanchet:LRR, FutamaseItoh:LRR} and \cite{Maggiore}.

The PN expansion is valid in a region,  called the near zone, smaller
than the wavelength of the emitted gravitational waves. Applying the
PN approximation far from the source (in the far zone) can lead to
logarithmic divergences since retardation effects from the finite
propagation speed of gravitational waves cannot be neglected far from
the sources. As a result, the PN metric in the near zone is matched
to a metric containing the gravitational radiation emitted by the
binary. This matching is performed in an overlapping region of the
near and far zones. In this way, one can describe the generation and
propagation of gravitational waves by slowly moving compact objects
in the PN framework \cite{Thorne:RevModPhys52, Damour:300yrs}.

The metric in the far zone can be calculated from  a post-Minkowski
(PM) expansion where the Einstein equations are expanded in powers of
Newton's constant $G_N$ and solved iteratively for each order of the
metric perturbations. In particular, there is no constraint on the
velocities of the sources. This method is valid in those regions of
spacetime that are weakly gravitating so the PM expansion is not
applicable for compact binaries everywhere in the spacetime. However,
as discussed above, the PM expansion can be applied far from such a
system by matching onto the perturbed metric of the near zone
computed in the PN approximation.

{The traditional approaches to the post-Newtonian expansion outlined
above have been efficiently streamlined by Goldberger and Rothstein
\cite{GoldbergerRothstein:PRD73} who borrow time-honored techniques
from quantum field theory. For example, Feynman diagrams are utilized
to construct the PN expansion order by order and the divergences
typically encountered in a theory of interacting fields and point
particles are conveniently regularized using dimensional
regularization \cite{tHooftVeltman:NuclPhysB44}. Furthermore, a
systematic formalism is introduced in
\cite{GoldbergerRothstein:PRD73} that allows for a point particle
description of an extended body by utilizing an effective field
theory paradigm and including all possible extra terms (so-called
non-minimal couplings) into the point particle Lagrangian that are
consistent with general coordinate invariance and reparamaterization
invariance.}

Recent years have seen the development of the so-called effective
one-body (EOB) approach \cite{BuonannoDamour:PRD59,
BuonannoDamour:PRD62} that uses the 3PN expanded Hamiltonian to resum
the conservative terms by mapping the two-body dynamics onto a system
describing a test particle moving in an effective background
spacetime. This background is a deformed Schwarzschild spacetime
where the deformation parameter is the symmetric mass ratio, $\nu = m
M / (m+M)^2$. By following the geodesic motion of the test particle
on this effective geometry one can describe the conservative dynamics
of the binary even in the relativistic regime. The effects of
radiation reaction are ``grafted" onto the conservative test particle
dynamics by using a flux balance argument where the power emitted in
gravitational waves equals the loss in the particle's mechanical
energy due to the reactive forces that come from emitting the
gravitational radiation.

While the EOB method  has proved remarkably successful for providing
approximate analytical waveforms of nearly equal mass non-spinning
binaries it has several shortcomings from both a practical and
theoretical point of view. On the practical side, the EOB approach
does not yet describe well the dynamics of spinning binaries.
Furthermore, for non-spinning binaries, one must patch the
inspiral/plunge phases to the merger and the merger to the ringdown
phase because the EOB approach does not smoothly describe these
transitions. On the theoretical side,
the EOB approach can be regarded as a phenomenological model with
parameters that {are obtained by fitting to numerical simulations of
inspiral waveforms, for example.}  See \cite{Boyle_etal:PRD78,
DamourNagar:0902.0136, Buonanno_etal:0902.0790} and references
therein for further discussions.

\subsubsection{Black hole perturbation theory for extreme mass ratios}

The extreme mass ratio inspiral (EMRI) scenario  consists of two
bodies with largely disparate masses. One body, a supermassive black
hole (SMBH), has a mass so much larger than the other, a small
compact object, that the dominant geometry of the spacetime is
determined by the SMBH. Despite having a very much smaller mass than
the first, the small compact object nevertheless perturbs the
background black hole spacetime and brings about the emission of
gravitational radiation which causes the smaller mass to eventually
spiral in toward the SMBH. Its backreaction on the small compact
object generates a radiation reaction force which is called the `self
force'. It is a result of gravitational waves back-scattering off of
the background spacetime and encountering the compact object at a
later time and place in its orbit. This intrinsically nonlocal
(dependent on the object's pass history) property is the main
challenge in the calculation of the self force in this class of
problems.  See \cite{Poisson:LRR} for a good review of this approach.

The perturbation theory for EMRIs has the advantage of treating the
relativistic motion of the small compact object in a strongly curved spacetime.
For detecting gravitational waves the space-based gravitational wave
observatory LISA only requires knowing the self-force
and the radiation through first order in the small mass ratio. For precise
parameter estimation the self-force and the gravitational radiation
will likely need to be calculated to second and third orders, respectively \cite{Burko:PRD67}.

\subsection{New analytic methods for arbitrary mass ratios and relative velocity}

{The approaches briefly reviewed above are largely useful for
describing either nearly equal mass binaries (PN) or extreme mass
ratio systems. While these constitute a large class of expected
observable gravitational wave signals for both ground-based and
space-based interferometer detectors there may also be signals coming
from compact binaries having generally unequal or intermediate mass
ratios. Even though the event rates of such systems are not
accurately known, intermediate mass ratio inspirals (IMRIs) may occur
in the dense centers of globular clusters \cite{FrankRees}.
Furthermore, during the final stages of galaxy mergers there may
potentially be detectable gravitational wave signals from the
inspiral of generally unequal SMBHs that originally formed the cores
of the merging galaxies. Indeed, the potential discovery of an
intermediate mass ratio binary \cite{BorosonLauer:Nature458} (see
also \cite{Gaskell:0903.4447} for an alternative interpretation of
the data) suggests that IMRI sources may be present throughout the
universe.}

Not much work has been done on IMRI systems, not because it is deemed
less important or urgent but usually considered a very difficult
task, as it falls in a hard- to- reach territory between the two
familiar terrains, namely, the PN expansion and black hole
perturbation theory. Simulating the dynamics of IMRI systems is not
yet possible with numerical methods. As the mass ratio decreases, the
computational cost and incurred numerical errors both increase
dramatically making it difficult and very expensive to numerically
evolve binaries with relatively different masses. For this reason
even with small gains it would still be worthwhile to explore new
analytic methods for inspiraling binaries having general mass ratios
and relative velocities.  Here we present two new analytical
techniques to two different parameter regimes of inspiraling binaries
with unequal masses, including IMRIs.

\subsubsection{Gravitational perturbation theory with self-consistent backreaction (GP-SCB)}

We have at least two motivations for inventing a new formalism. The
first comes from a practical aim: to provide a common framework
encompassing these two major schemes - the (equal mass)
post-Newtonian scheme and the (extreme mass ratio) perturbation
theory, but extending the limitations from both ends (of the mass
ratio parameter). We want this framework to be applicable to
intermediate mass ratio processes without the slow motion, weak field
restrictions.

The second motivation comes from a more idealistic goal: to build a
fully self-consistent formalism of perturbation  theory that is able
to account for the backreaction {\it from all} of the dynamical
variables and {\it on all} of them. These variables (in a
perturbative context here) are the motion of the compact objects
(which is affected by radiation reaction and self-force
from gravitational wave emission and by the tidal forces each object
exerts on the other), the gravitational waves emitted by the binary,
and the background metric (which together with the gravitational
waves give the full metric). The full metric is dynamically
determined by the negotiation between the two compact objects
{through mediation} by the gravitational waves.

To aid in such an effort we describe a generic framework  for
generating perturbation theories that is self-consistent in the sense
that the participating dynamical quantities are conserved
(in each perturbative order)
on the background spacetime and have the appropriate behavior under
general (i.e., not necessarily infinitesimal or small) coordinate
transformations. This framework is based on previous work of Anderson
\cite{Anderson:PRD55} for studying the self-consistency of the
gravitational geon solution in \cite{AndersonBrill:PRD56}. As a
useful check of this new formalism we demonstrate how to obtain the
well-established post-Newtonian expansion and the extreme mass ratio
perturbation theory as its subcases.

To apply this new formalism it helps if one can identify some
physical parameter  in certain stages of the binary's motion with a
clear discrepancy in the mass, velocity or frequency most suitable
for a perturbative expansion \footnote{Broader examples include the
two-time (slow-fast) expansion in statistical physics, rotating wave
approximation in atomic physics, Born-Oppenheimer approximation in
molecular physics, mass hierarchy and effective cutoffs in particle
physics}.
As an example, we consider those cases where the dynamics of the
binary includes the existence of an inspiral phase. An inspiraling
binary evolves with an orbital period that is shorter than the
secular time scale associated with the radiation reaction and
dissipative self-forces. We demonstrate that averaging over the fast
orbital period leads to a perturbation theory where the mutual
backreaction from and on all of the dynamical variables involved are
incorporated. In this case the companion-induced tidal moments are
included into the description of the compact object(s)
\cite{GoldbergerRothstein:PRD73}. Because this  perturbation theory
does not depend on the binary's mass ratio or relative velocity, at
least explicitly for its formal construction, we have reasons to
believe that our treatment in this example may be applicable to
inspiraling binaries with generic masses and velocities, including
intermediate mass ratio binaries and comparable mass binaries, and
possibly to the plunge and merger phases
because of the inclusion of backreaction effects. Additional
assumptions and approximations may be useful for practical
calculations but they are not necessary for the internal consistency
of this new theory.

The self-consistency requirement is a crucial structural feature of
this new approach, and the allowance for dynamical backreaction is an
attractive functional feature because the background can respond to
the effective stresses and energies arising from the motion of the
compact objects and the emitted gravitational waves.  This is to be
contrasted with other prevailing approaches, foremost the PN and EMRI
perturbative schemes, that choose a fixed background which never
deviates from its originally specified form. While this is clearly
convenient for calculations of many scenarios, especially if the
fixed background possesses some isometries,  it is not general enough
for a wider class of binary processes and not very realistic on a
broader scope, because of the nonlinear interactions between the
massive objects and the nonlocal (history dependent) influences of
the gravitational waves.

\subsubsection{Post-Newtonian effective field theory (PN-EFT) for
unequal masses}

The PN expansion is often regarded as being applicable to nearly
equal mass binaries or to the weak-field evolution of an extreme mass
ratio binary. However, many interesting gravitational wave signals
for LIGO and LISA come from evolution in a strong field regime,
including the relativistic motion of a compact object near a
supermassive black hole and the plunge and merger phases of an
astrophysical equal mass binary.

To develop equations of motion that are based on a
PN approximation but can also be applied to unequal mass binaries in
stronger fields requires introducing the mass ratio $\eta = m/M$ as
an additional expansion parameter to the velocity $v$ and developing
the current PN expressions to higher orders. Within the context of
the effective field theory approach originally developed by
Goldberger and Rothstein \cite{GoldbergerRothstein:PRD73}, who called
this approach `non-relativistic general relativity' (NRGR)
\footnote{This nomenclature seems natural in the context where it is
adopted from, referring to the nonrelativistic limit of QCD \cite{LukeManoharRothstein:PRD61}. However,
it could be a bit confusing when transposed to general relativity,
having `Nonrelativistic relativity' in one phrase where
`relativistic' refers to special relativity (slow motion) and
`relativity' refers to general relativity (GR). In the GR community
the term post-Newtonian (PN) has been used for decades and already
encompasses the nonrelativistic meaning in NRGR. To us the
introduction of effective field theory (EFT) concepts and
techniques to GR is a key contribution by these authors and hence we
suggest calling this theory post-Newtonian effective field theory or
PN-EFT.}, we can include $\eta$  into the power counting scheme for
determining the scaling of the interaction terms that
contribute to the PN expansion at a given order in $v$ {\it and}
$\eta$. In fact, in the context of PN-EFT, this can be done systematically and efficiently to
any order.

For a given mass ratio and total mass it is instructive to have some
idea of the  PN order that a quantity, such as the number of cycles
of a gravitational waveform that fall into LIGO's bandwidth, should
be expanded through. With such estimates one can begin to more
accurately compute PN expanded equations of motion, waveforms, etc.
for binaries that have mass ratios of $10^{-1}$ or $10^{-2}$, for
example. For such systems, the smaller compact object can be
accelerated to higher speeds and so a knowledge of higher order PN
terms becomes important for accurately describing the binary.

\subsection{Organization}

This paper is organized as follows.  In Section \ref{sec:bs} we
present this new perturbation theory for gravitational binary systems
with self-consistent backreaction from and on all dynamical variables
involved  within a rather general framework, viz., the compact object
motion, its tidal deformations, gravitational waves, {\it and} the
background spacetime geometry. We demonstrate how the post-Newtonian
expansion and the perturbation theory for describing EMRIs arise from
within this formalism. As an example, we apply this new formalism to
dynamics which has an inspiral phase for any mass ratio and relative
velocity.  In Section \ref{sec:pneftunequal} we generalize the PN-EFT
approach to include the mass ratio as an expansion parameter. We then
provide crude estimates that indicate the post-Newtonian orders
needed, with a given precision, to calculate the number of cycles
observed in LIGO's bandwidth for a binary with general values for its
masses. We end with a discussion of directions for future
development. Further details for each of these approaches are
contained in two forthcoming papers \cite{GalleyHu:SCB1, GalleyHu:SCB2}.

\section{Gravitational perturbation theory with self-consistent backreaction}
\label{sec:bs}

Our goal is to establish a formalism which describes the evolution of
three dynamical variables in a self-consistent manner: the spacetime
and its perturbations (the gravitational waves), and the motion of
the compact objects. We take a first principles approach that is
sufficiently flexible and general to accommodate several
approximation schemes and perturbation theories. This general theory
should encompass all the existing approaches as delineated in the
last section yet be able to treat parameter ranges outside of their
validity. There are a few basic criteria we set for any such
formalism to obey, foremost a gauge-invariant effective stress-energy
tensor for gravitational waves. For these purposes we adopt the
framework of Anderson  \cite{Anderson:PRD55} who provided a rigorous
foundation for the Brill-Hartle-Isaacson averaging procedure
\cite{BrillHartle:PhysRev135, Isaacson:PhysRev166_1,
Isaacson:PhysRev166_2} and applied it to a careful study of
gravitational geons \cite{AndersonBrill:PRD56}.

The two-body problem may be described at the level of the equations
of motion in several ways, depending on the physical setup, the
processes, and what specific parameter values are of interest. {There
are three classes of compact binaries (i.e., radiating binary systems
composed of compact objects) typically regarded in the literature: 1)
black hole/black hole (BH/BH), 2) black hole/neutron star (BH/NS),
and 3) neutron star/neutron star (NS/NS). }

{In a first principles description, the spacetime geometry of a BH/BH
binary is described by the vacuum Einstein equations
\begin{equation}
    G_{\mu\nu} (g) = 0
    \label{EE1}
\end{equation}
where $g_{\mu\nu}$ is the full metric of the spacetime. When a
binary is composed of one or two neutron stars one can describe the
system in terms of a matter stress-energy via the non-vacuum Einstein
equations,
\begin{equation}
    G_{\mu\nu} (g) = \sum 8 \pi G_N T_{\mu\nu}^{NS} (g; \rho, p, \ldots)
    \label{EE2}
\end{equation}
where the summation is included depending on  whether one is
considering BH/NS or NS/NS binaries. Here $g_{\mu\nu}$ is the full
spacetime metric, which is determined by the dynamics of the binary,
and $T_{\mu\nu}^{NS}$ is the stress-energy tensor appropriate for a
neutron star with density $\rho$, pressure $p$, etc. At present, the
dependence of this stress tensor on the fluid variables is not well
understood because the NS equation of state is not known. }

{Solving (\ref{EE1}) or (\ref{EE2}) can only be done using numerical
techniques. Even then, most simulations are carried out for nearly
equal mass binary black holes (i.e., mass ratios $\gsim 1/4$) and
numerical methods for BH/NS and NS/NS binaries are still in their
infancy.}

{Analytically solving (\ref{EE1}) or (\ref{EE2}) can be accomplished
by invoking approximations and restricting solutions to a particular
dynamical regime (e.g., inspiral or ringdown phases). In a BH/BH
binary having one mass considerably smaller than the other, as in the
extreme mass ratio scenario, one can model the smaller BH using a
point particle description. In this case, the Einstein equation
(\ref{EE1}) is approximately
\begin{equation}
    G_{\mu\nu} (g) = 8 \pi G_N T_{\mu\nu}^{pp} (g; z)
    \label{EE3}
\end{equation}
where the coordinates of the particle worldline  are denoted by
$z^\mu$. Here $g_{\mu\nu}$ is (close to but not quite exactly)
the full metric of the binary spacetime (as obtained from \ref{EE1}).
For very small mass ratios  one could provide an equivalent
description to a high degree of accuracy by replacing the binary
system by a background spacetime generated by the larger BH otherwise
in isolation plus a perturbation generated by the presence and motion
of the smaller BH. As such, one may decompose the full metric into a
fixed background $g_{0\mu\nu}$ (e.g., describing a Kerr black hole)
and its perturbations $h_{\mu\nu}$. By expanding (\ref{EE3}) in
powers of $h_{\mu\nu}$ and the mass ratio one can perturbatively
obtain analytic solutions for a binary BH in the extreme mass ratio
limit.}

{From the considerations above, the equations of motion for a general
compact binary system can be generically written as
\begin{eqnarray}
    \cG_{\mu\nu} (g, \alpha) = 0
    \label{EEfull0}
\end{eqnarray}
where
\begin{equation}
    \cG_{\mu\nu} (g , \alpha) \equiv G_{\mu\nu} (g) - \cT_{\mu\nu} (g, \alpha)
    \label{EEfull1}
\end{equation}
and $\cT_{\mu\nu}$ is proportional to the stress-energy for the
compact object(s),  if the assumptions and input allows for the
inclusion of material stress-energy. In (\ref{EEfull1}), $\alpha$
denotes the collection of variables that describe the dynamics of the
compact object(s), such as the density, pressure, worldline
coordinates, etc. The number of contributing stress tensors to
$\cT_{\mu\nu}$ and the forms of these contributions depend crucially
on how one describes the constituents of the binary (e.g., as a point
particle, as an extended fluid body), the regime of dynamical
evolution being considered (e.g., inspiral, ringdown), etc.
 Below we closely follow the
presentation given in \cite{Anderson:PRD55}.}

\subsection{Generalized gauge transformations, gauge invariance, and self-consistency}

In an analytic approach the metric of the full spacetime $g_{\mu\nu}$
is separated into a background part $g_{0\mu\nu}$ and a perturbation
part $h_{\mu\nu}$ so that $g_{\mu\nu} = g_{0\mu\nu} + h_{\mu\nu}$. We
stress that this decomposition is completely arbitrary. Indeed, in
standard perturbation expansions one fixes the background metric
according to the problem at hand (e.g. Kerr background for EMRIs).
The unperturbed Einstein equation (\ref{EEfull0}) can generally be
written according to the above metric decomposition in the form
\footnote{We will often drop the spacetime indices for notational
convenience and will suppress the dependence on the background metric
and worldline for perturbed quantities.}
\begin{equation}
    \cG_{\mu\nu} (g_0, \alpha) + \Delta \cG_{\mu\nu} ( h,\alpha) = 0
    \label{EEpert0}
\end{equation}
or equivalently as
\begin{equation}
    G (g_{0}) + \Delta G ( h) = \cT (g_0, \alpha) + \Delta \cT (h, \alpha)
    \label{EEpert1}
\end{equation}
Because we want the metric decomposition to be arbitrary the quantities
$\Delta G_{\mu\nu}$ and $\Delta \cT_{\mu\nu}$ contain all of the
dependence on the gravitational perturbations $h_{\mu\nu}$ and are
not necessarily small with respect to the background Einstein or
stress tensors, $G_{\mu\nu}(g_{0})$ and $\cT_{\mu\nu} (g_{0}, \alpha)$,
respectively.

For the formalism to be self-consistent we require that the perturbed quantity
\begin{equation}
    \Delta \cG ( h, \alpha) = \Delta G ( h) -  \Delta \cT ( h, \alpha)
    \label{pert0}
\end{equation}
is both conserved and invariant under coordinate transformations that preserve the structure of the background geometry. The former is straightforward to demonstrate.

Given a solution $(g_{0\mu\nu} , \alpha )$ to (\ref{EEpert1})
we know that $G(g_{0})$ and $\cT (g_{0}, \alpha)$ are separately
conserved with respect to the background geometry upon using the
Bianchi identities for the Einstein tensor and the conservation
equation for the stress tensor, which gives the
equations of motion for the compact object variables on the background geometry. It therefore
follows from (\ref{EEpert1}) that  (\ref{pert0}) is also conserved
with respect to the background geometry, $\nabla^\mu \Delta \cG_{\mu\nu} ( h,
\alpha) = 0$ where $\nabla_\mu$ is the covariant derivative compatible
with $g_{0\mu\nu}$.

Since the perturbed quantity $\Delta \cG ( h, \alpha)$ is not necessarily small with respect to
$\cG (g_{0})$ it follows that the coordinate transformations, which change the perturbed metric but not
the background, are
not necessarily infinitesimal as is usually considered for
discussions of gauge invariance. As such these coordinate
transformations are called {\it generalized} gauge transformations
\cite{Anderson:PRD55}.

A coordinate transformation can always be written in the following form
\begin{equation}
    \barx^\mu = x^\mu + \xi^\mu   \label{coordtrans0}
\end{equation}
where $\xi^\mu$ is a function of $x^\mu$ and is not necessarily infinitesimal nor small. Accordingly, the full metric changes via the usual tensor transformation rule
so that the metric perturbations in the two coordinates are related through
\begin{align}
    g_{0\mu\nu} (x) + h_{\mu\nu} (x) ={}& g_{0\mu\nu} (\barx) + \barh_{\mu\nu} (\barx) +
    \big[ g_{0\mu \alpha} (\barx) + \barh_{\mu\alpha} (\barx) \big] \xi^\alpha_{ ~ ,\nu} + \big[ g_{0\alpha \nu} (\barx) + \barh_{\alpha \nu} (\barx) \big] \xi^\alpha _{~,\mu} \nonumber \\
    &+ \big[ g_{0\alpha \beta} (\barx) + \barh_{\alpha \beta} (\barx) \big] \xi^\alpha_{~,\mu} \xi^\beta_{~,\nu}
    \label{coordtrans1}
\end{align}
where the derivatives of $\xi^\mu$ are with respect to $x^\mu$. In the limit of infinitesimal $\xi^\mu$ we recover the usual form for the infinitesimal coordinate transformation of the metric perturbations
\begin{equation}
    \barh_{\mu\nu} (\barx) = h_{\mu\nu} (x) - g_{0\mu\nu , \alpha} (x) \xi^\alpha - g_{0\mu \alpha} (x)
    \xi^\alpha _{~, \nu} - g_{0\alpha \nu} (x) \xi^\alpha _{~, \mu}
\end{equation}
The perturbed quantity $\Delta \cG (h, \alpha)$ is gauge invariant under a generalized gauge transformation (\ref{coordtrans0}) if the following relation holds
\begin{equation}
    \Delta \cG ( \barh , \alpha) = \Delta \cG (h, \alpha)
\end{equation}
where $\barh_{\mu\nu}$ is found by inverting (\ref{coordtrans1}). The
proof can be found in \cite{Anderson:PRD55}.

So far we have described a general framework that is self-consistent
in the sense that $\Delta \cG$ is both conserved on the background
geometry and is gauge-invariant under coordinate transformations of
the form (\ref{coordtrans0}). However, we cannot solve for the
background or the perturbations from the Einstein equations
\begin{equation}
    \cG (g_{0} , \alpha) = - \Delta \cG (h, \alpha)
    \label{EEpert2}
\end{equation}
alone without knowing how the full spacetime geometry is decomposed
into a background and its perturbation. There are at least three ways
to do this, in principle. The first is to specify the background and
solve for the perturbation or specify the perturbation and solve for
the background.

The second is to define a conserved and gauge invariant effective
stress energy tensor (or in some approximate sense) for gravitational
waves $T^{gw}_{\mu\nu}$ on the background so that
\begin{align}
    \Delta \cG (h , \alpha) &= -8 \pi G_N T^{gw} ( h, \alpha )  \\
    \cG (g_{0} , \alpha) &= 8 \pi G_N T^{gw} ( h, \alpha)
\end{align}
For example, one could define
\begin{equation}
    T^{gw} ( h, \alpha) = - \frac{1}{8\pi G_N} \big\langle \Delta \cG ( h , \alpha ) \big\rangle
\end{equation}
which is clearly gauge invariant under coordinate transformations of
the form (\ref{coordtrans0}). Here the angled brackets denote a time,
space, or space-time average. We note that this stress tensor is
related to the choice made by Brill and Hartle  in
\cite{BrillHartle:PhysRev135} with time averaging and by Isaacson
\cite{Isaacson:PhysRev166_1, Isaacson:PhysRev166_2} with space-time
averaging.

The third is to introduce a gauge-invariant equation that  determines
the metric perturbation thereby fixing the spacetime decomposition.
For gravitational waves this is naturally provided by a wave equation
\begin{equation}
    H (g_{0}, h , \alpha ) = 0
\end{equation}
We will discuss this approach in more detail below with a couple of
examples.

In most situations encountered in the two-body problem the  equations
cannot be solved exactly and some approximations must be invoked,
usually in the form of perturbation expansions. Let us assume that
there exists an expansion of $\Delta \cG$ so that
\begin{equation}
    \Delta \cG = \Delta_1 \cG + \Delta _2 \cG + \cdots
    \label{expandPert0}
\end{equation}
Here the subscript on a $\Delta$ denotes the order of the expansion.
We remark that we have not actually specified any particular
assumptions regarding the expansion such as the strength of the
gravitational waves or its derivatives nor have we identified a small
parameter to do the expansion. We merely assume that there exists a
valid expansion of the form (\ref{expandPert0}). As such, it is not
necessarily true that $\Delta_2 \cG$ is strictly quadratic in $h$,
for example, as is the case for the gravitational geon
\cite{BrillHartle:PhysRev135, Anderson:PRD55}.

If we truncate the expansion in (\ref{expandPert0}) to order $n$ then
the discussions above imply that $\Delta_1 \cG + \cdots + \Delta_n
\cG$ is conserved with respect to the background geometry to $n^{\rm
th}$ order. Gauge invariance of (\ref{expandPert0}) can also be shown
provided that the generalized gauge transformations are such that
$\barh$ and $h$ are of the same order of magnitude to facilitate
comparison  \cite{Anderson:PRD55, Isaacson:PhysRev166_1,
Isaacson:PhysRev166_2}, which we will assume from here on. It is
important to realize that the quantity $\Delta_1 \cG + \cdots +
\Delta_n \cG$ is gauge invariant to $n^{\rm th}$ order. Therefore,
$\Delta_1 \cG$ is gauge invariant to first order but $\Delta_2 \cG$
is generally not gauge invariant at any order. Instead it is the sum
$\Delta_1 \cG + \Delta_2 \cG$ that is gauge invariant to second
order.

\subsection{Existing theories as special cases}
\label{sec:specialcases}

In this section we will show that the general perturbative framework
developed above contains the familiar existing theories including 1)
the (Regge-Wheeler) perturbation theory for EMRIs and 2) the
post-Newtonian expansion for equal mass binaries. In the following
section we will generate a new expansion that may be largely
insensitive  to the mass ratio and the typical velocities of the
binary.

\subsubsection{Perturbation theory for EMRIs}

To build a perturbation theory for EMRIs, use the smallness of the
binary's mass ratio $\epsilon= m/M \ll 1$ so that $h_{\mu\nu} =
O(\epsilon)$ and $T^{pp} (g_{0}, z) = O(\epsilon)$ where $T^{pp}$ is the stress tensor of a point particle, which models the small compact object. Notice that we
are not saying anything yet regarding the nature of the background
spacetime, namely, whether it is fixed or dynamical, etc. Expand the Einstein equations after decomposing the
full metric as $g= g_{0} + h$ to find
\begin{equation}
    G( g_{0}) + \Delta G( h) = \Delta \cT ( h, z) = 8 \pi G_N T^{pp} (g_0, z) + 8 \pi G_N \Delta T^{pp} (h, z)
    \label{emrifull0}
\end{equation}
Notice that the source of curvature should be regarded as a perturbed quantity since the point particle stress tensor is proportional to the small mass $m$ of the compact object.
Expanding the perturbed quantities $\Delta G$ and $\Delta T$ in powers of $\epsilon$ gives to second order
\begin{align}
    \Delta G( h) & = \Delta_1 G(h) + \Delta_2 G(h) + \cdots \\
    \Delta \cT( h, z) &= \Delta_1 \cT( z) + \Delta_2 \cT(  h, z) + \cdots
\end{align}
where $\Delta_{1,2}$ denotes a quantity of $O(\epsilon^{1,2})$. Because the particle stress tensor is already proportional to $m$ then $\Delta_1 \cT(z)$ is independent of $h$.

We can fix the metric decomposition by specifying a gauge invariant
but otherwise arbitrary equation for the perturbations, $H(g_{0}, h
,z)=0$. A natural choice for $H$ is a wave equation such that
\begin{equation}
    H (g_{0}, h, z) = \Delta G( h) - \Delta \cT(h, z) = 0
    \label{wave0}
\end{equation}
From (\ref{wave0}) and (\ref{emrifull0}) it follows that the leading order contribution to the Einstein equations is
\begin{equation}
    G(g_{0} ) = 0 \label{emribg0}
\end{equation}
Note that according to (\ref{emribg0}) the background geometry is not affected by the gravitational waves. Indeed, the background metric describes a vacuum spacetime that one can fix or specify {\it ab initio}.
To solve the wave equation order by order in $\epsilon$ we write
\begin{equation}
    h = h_1 + h_2 + O(\epsilon^3)
    \label{expandhemri0}
\end{equation}
so that (\ref{wave0}) becomes
\begin{gather}
    \Delta_1 G(h_1) =  \Delta_1 \cT( z)  \\
    \Delta_1 G(h_2) = - \Delta_2 G(h_1)+  \Delta_2 \cT (h_1,z)
\end{gather}
through second order in $\epsilon$. The first line simply describes the wave equation for the first order perturbation $h_1$, which is sourced by the stress tensor for the point particle with worldline coordinates $z^\mu$.

The particle's equations of motion follow from the geodesic equation
$a^\mu(g, z) = 0$ due to the conservation of $\Delta \cT$ on the full
spacetime $g$. Using the metric
decomposition $g = g_{0} + h$ and the expansion of $h$ in
(\ref{expandhemri0}) we see that through first order
\begin{equation}
    a(g_0, z) = - \Delta_1 a(h_1, z) ,
\end{equation}
which can be written in a more recognizable form as
\begin{equation}
    u^\alpha (\tau) \nabla_\alpha u^\mu (\tau) = - \left( \nabla_\alpha h^{~\mu} _{1 ~ \beta} (z) - \frac{ 1}{2} \nabla^\mu h _{1 \, \alpha \beta} (z) \right) u^\alpha (\tau) u^\beta (\tau)
\end{equation}
where $\tau$ is the proper time for the worldline $z^\mu$ and $u^\mu = dz^\mu/d\tau$. Indeed, this is the unregularized equation describing the first order self-force on the small compact object, which was first derived (and regularized) in \cite{MinoSasakiTanaka:PRD55, QuinnWald:PRD56}.

\subsubsection{Post-Minkowski and post-Newtonian expansions}

A post-Minkowski expansion is built by including a stress tensor for each of the two compact objects and choosing Newton's constant $G_N$ as the expansion parameter \footnote{More precisely, the condition is that $G_N m/r$ and $G_N M/r$ are both small throughout the spacetime.}. We find the following self-consistent perturbed field equations in a manner similar to the previous section
\begin{gather}
    G(g_{0} )  = 0  \\
    \Delta_1 G(h_1)  = \Delta_1 \cT (z, Z) \\
    \Delta_1 G(h_2)  = - \Delta_2 G( h_1) + \Delta_2 \cT (h_1, z, Z)
\end{gather}
where $z^\mu$ and $Z^\mu$
denote the worldline coordinates of the masses $m$ and $M$,
respectively, and
\begin{align}
    \Delta_1 \cT (z, Z) & = 8 \pi G_N T^{pp}_{(m)} (g_0, z) + 8 \pi G_N T^{pp} _{(M)} (g_0, Z) \\
    \Delta_2 \cT (h_1, z, Z) & = 8 \pi G_N \Delta_1 T^{pp} _{(m)} (h_1, z) + 8 \pi G_N \Delta_1 T^{pp} _{(M)} (h_1, Z)
\end{align}
Again, the leading order equations indicate that the
background is vacuous and the considerations of the specific problem
imply $g_{0\mu\nu} = \eta_{\mu\nu}$.

In the near-zone region of a comparable mass binary, where the gravitational perturbations propagate nearly instantaneously, the compact objects move slowly compared to the speed of light. Consequently, we use the relative velocity $v$ as an expansion parameter. Together with the post-Minkowski expansion (where we assume that $G_N = O(v^2)$ from the virial theorem) we can construct a valid perturbation theory of the Einstein equations and particle equations of motion.

Following the similar steps as in the previous section we find at
leading order that $G(g_{0}) = 0$, which indicates that the background
can be fixed to Minkowski spacetime as before and that the gauge
invariant wave equation is
\begin{equation}
    \Delta G(h, z) = \Delta \cT (h, z, Z)
\end{equation}
At first order in $v$ we find the perturbed wave equation is $\Delta_1 G(h_1) = 0$ where the right hand side is zero because the leading order contribution from each point particle stress tensor is $O(G_N) = O(v^2)$, which is a second order quantity. Therefore, we can consistently choose $h_1=0$. At second order in $v$ the perturbed wave equation is
\begin{equation}
    \Delta_1 G(h_2) =  \Delta_2 \cT (h, z, Z) = 8 \pi G_N T_{(m)} (\eta, z)  + 8 \pi G_N T_{(M)} (\eta, Z)
    \label{secondorderpn0}
\end{equation}
where we have used that $h_1 =0$. One can easily show that when a gauge is chosen for the metric perturbations (e.g., the harmonic gauge on a flat background) that (\ref{secondorderpn0}) is just the Poisson equation for $h_{00}$ sourced by the two point particles on worldlines with coordinates $z^\mu$ and $Z^\mu$.

The particle equations of motion follow from the geodesic equation in the full spacetime so that through $O(v^2)$ we find
\begin{gather}
    a_{(m)}(\eta, z) = - \Delta_1 a_{(m)} (h_2, z) \\
    a_{(M)} (\eta, Z) = - \Delta_1 a_{(M)} (h_2, Z)
\end{gather}
which are the equations for two particles moving in their respective Newtonian gravitational potentials.
Continuing to higher orders merely reproduces the equations of motion for the field and the particles at successively higher post-Newtonian orders.

\subsection{Two-time separation and adiabatic expansion in the GP-SCB formalism applied to the inspiral phase}
\label{sec:adiabaticbkrxn}

In this section we illustrate how the general perturbation formalism
with self-consistent backreaction can be applied to a class of
problems with the introduction of a two-time separation
scheme. Throughout this section we consider the motion of a compact object with mass $m$ (a
neutron star or black hole) in the spacetime of a larger black hole
with mass $M>m$.
While it is convenient to use a point particle description for the
smaller compact object it is difficult to justify such an
approximation for an extended body. Even if it is acceptable for a
certain period of time one cannot assume this is valid throughout the
entire course of its evolution for binaries with arbitrary mass
ratios because the larger mass will induce tidal effects that
manifest as higher order multipole moments for the distorted shape of
the smaller compact object. Such effects are not negligible during the plunge and merger phases, especially as $m \to M$.

To include the tidal effects in such binaries one needs to take into
account the backreaction effects. For comparable mass binaries the
compact objects are more severely disrupted during the inspiral, plunge
and/or merger stages in which case a point particle (pp)
description for the masses is more likely to become invalid. This is where
an effective point particle (epp) description beyond the pp approximation will prove useful in the GP-SCB scheme.

The non-minimal couplings that appear in the effective point particle description of an extended body \cite{GoldbergerRothstein:PRD73}
are related to terms in a multipole expansion that scale as powers of $r_m /
\lambda$ where $r_m$ is the size of the compact object with mass $m$
and $\lambda$ is the typical wavelength of an external gravitational
perturbation at some scale. For such a compact object plunging toward merger with another mass $M$,
\begin{equation}
    \frac{ r_m }{ \lambda } \sim \frac{ r_m }{ r } v \lsim \frac{ r_m}{r}
\end{equation}
for arbitrary velocities (The second relation comes from the
fact that for both EMRI and equal mass binaries, $v \approx 0.5 -
1.0$ near merger). Near the merger phase the orbital separation is
of order $r \sim r_m+r_M$ and it follows that
\begin{equation}
    \frac{r_m}{ \lambda } \lsim \frac{ m }{ m +M }
\end{equation}
Thus if the mass ratio is $\lsim 1$ then one can justify using a
point particle description for the smaller mass so long as a
sufficient number of non-minimal terms (possibly many) are included
to account for the companion-induced tidal moments.

The action describing the effective point particle (epp) dynamics for the small compact object is given by \cite{GoldbergerRothstein:PRD73}
\begin{align}
    S_{epp} [ z, g] = & - m \int d\tau + c_R \int d\tau \, R(z) + c_V \int d\tau \, R_{\alpha \beta} (z) \dz^\alpha \dz^\beta \nonumber \\
    & + c_E \int d\tau \, \cE_{\alpha \beta} (z) \cE^{\alpha \beta} (z) + c_B \int d\tau \, \cB_{\alpha \beta} (z) \cB^{\alpha \beta} (z) + \cdots
    \label{effppaction0}
\end{align}
where the quantities $\cE_{\alpha \beta}$ and $\cB_{\alpha \beta}$
are the electric and magnetic parts, respectively, of the Weyl
curvature tensor. The constants $\{c_i\}$ depend on the internal
structure of the compact object (e.g., parameters of the equation of
state for a neutron star) and can be calculated by matching onto a
full description of the ``microscopic" theory for the compact object.
See \cite{GoldbergerRothstein:PRD73} for further discussion. From
this effective point particle action one can obtain the stress tensor
for the smaller compact object
\begin{equation}
    T_{\mu\nu}^{epp} (g, z) = \frac{2}{g^{1/2}}  \frac{ \delta S_{epp} }{ \delta g^{\mu\nu} }
\end{equation}
from which we obtain the Einstein equation for the binary
\begin{equation}
    G_{\mu\nu} (g) = 8 \pi T_{\mu\nu} ^{epp} (g, z)
    \label{EEexample0}
\end{equation}
Because of the non-minimal couplings in (\ref{effppaction0}) the
motion of the compact object $m$ is not given by the geodesic
equation. In fact, the acceleration of the smaller mass $m$ moving in
the full spacetime is given by the conservation of $T^{epp}$ in the full spacetime,
\begin{equation}
    \big( m - c_R R(z) + \cdots \big) a^\mu (\tau) = c_R \big( g^{\mu\nu} + \dz^\mu \dz^\nu \big) \nabla_\nu R(z) + \cdots  ,
    \label{ppeom0}
\end{equation}
and is non-zero precisely because the compact object is an extended
body although it is being modeled with a point particle
description. Notice the appearance of an effective mass resulting
from these non-minimal interactions \footnote{In a spacetime that is
vacuum to leading order in an expansion one can remove all of the
Ricci tensors in (\ref{effppaction0}); see
\cite{GoldbergerRothstein:PRD73} for elaboration of this point. We
will not do such a reduction here since we will see in (\ref{bggraveom0}) that the
background in our example is in fact not vacuous at leading order.}.
We can write the particle equations of motion (\ref{ppeom0}) in a
general form
\begin{equation}
    m^{\mu\nu} (g, z) a_\nu (\tau) = F^\mu (g, z; c_R, c_V, \ldots)
    \label{ppeom1}
\end{equation}
where the effective mass is a tensor that is generally time and space dependent, and $F^\mu$ accounts for the forces on the particle arising from the finite size effects of the tidally distorted compact object $m$. When such companion-induced tidal moments can be ignored (\ref{ppeom1}) reduces to the geodesic equation on the full spacetime with metric $g$.

Binaries exhibiting an inspiraling phase admit  the characteristic
feature wherein the orbital period $T_{orb}$ is often much shorter
than the timescale for radiation reaction and dissipative self-force
effects $T_{RR}$. Therefore, let us introduce an averaging procedure
that effectively averages the fast motion of the binary.

Split the spacetime into a background and its perturbation so that $g
= g_{0} + h$, and write the worldline coordinates of the effective
point particle as $z = z_0 + \zeta$ where $z_0$ denotes the
coordinates of the ``background" worldline and $\zeta$ its
perturbations. Then, introducing $\epsilon$ as the expansion
parameter, defined as the ratio $T_{orb}/ T_{RR}$, which is small
during at least the inspiral phase, we can expand (\ref{EEexample0})
and (\ref{ppeom1}) in a series giving
\begin{gather}
    G (g_{0}) + \Delta G( h)  = 8 \pi T^{epp} (g_{0}, z_0) + 8 \pi \Delta T^{epp} ( h, \zeta) \label{EEexample1}  \\
    m (g_{0}, z_0) a(g_{0}, z_0) + m(g_{0}, z_0) \Delta a( h, \zeta) + \Delta m( h, \zeta) a(g_{0}, z_0) + \Delta m (h, \zeta) \Delta a(h, \zeta)  = F (g_{0}, z_0) + \Delta F ( h, \zeta)
    \label{ppexample1}
\end{gather}
where we have dropped the spacetime indices in the particle equations of motion for notational simplicity.

Expand the perturbed quantities $\Delta G$ and $\Delta T^{epp}$ to
second order
\begin{align}
    \Delta G & = \Delta_1 G + \Delta _2 G + O(\epsilon^3) \\
    \Delta T^{epp} & = \Delta_1 T^{epp} + \Delta _2 T^{epp} + O(\epsilon^3)  .
\end{align}
We fix the metric decomposition by imposing a condition on the metric perturbations that uses an averaging procedure
\begin{align}
    \big\langle \Delta_1 G ( h) - 8 \pi G_N  \Delta_1 T^{epp} (h, \zeta) \big\rangle =
    0.
    \label{gaugecond0}
\end{align}
We have in mind that the average is over an orbit of the evolution. In particular, the average should be defined so that the secular decrease in the orbital period is incorporated in its definition. However, the precise definition of the average used here is not necessary for formulating the perturbation theory.

Notice that the condition in (\ref{gaugecond0}) is not exactly gauge invariant, which is simply a manifestation of the fact that fixing any metric decomposition is an intrinsically gauge-dependent procedure.
As such, (\ref{gaugecond0}) corresponds to choosing a particular gauge and so
will not necessarily remain invariant under a generalized gauge
transformation of the form (\ref{coordtrans0}). Nevertheless, the
quantity
\begin{equation}
    \big\langle \Delta_1 G (h) + \Delta_2 G(h) - 8 \pi G_N \Delta_1 T^{epp} (h, \zeta ) - 8 \pi G_N \Delta_2 T^{epp} (h, \zeta) \big\rangle
\end{equation}
will be gauge invariant through second order, as mentioned in earlier
discussions.
The condition (\ref{gaugecond0}) is a reasonable choice since the
fast and slow time scales of the orbital motion are imprinted into
the gravitational variables. In a sense, we are assuming that the
slow variation of the gravitational field ($\sim T_{RR}$) is
contained in the background $g_{0}$ while the relatively fast variations ($\sim
T_{orb}$) are contained in the metric perturbations $h$.

Further expanding the metric perturbation as
\begin{align}
    h & = h_1 + h_2 +O(\epsilon^3)
    \label{expandh0}
\end{align}
gives for (\ref{gaugecond0}) to second order
\begin{gather}
    \Delta_1 G ( h_2 ) \sim \Delta_2 G( h_1) \\
    \Delta_1 T^{epp} ( h_2, \zeta) \sim \Delta_2 T^{epp} (h_1, \zeta)
\end{gather}
where $\sim$ is to be read ``is of the same order as."
Using (\ref{expandh0}) we can
expand (\ref{gaugecond0}) in $\epsilon$ to find, order by order,
\begin{gather}
        \big\langle \Delta_1 G( h_1) - 8 \pi G_N \Delta_1 T^{epp} ( h_1,\zeta) \big\rangle = 0
        \label{fixmetric1} \\
        \big\langle \Delta_1 G( h_2) - 8 \pi G_N \Delta_1 T^{epp} ( h_2, \zeta)  \big\rangle = 0
        \label{fixmetric2}
\end{gather}
through $O(\epsilon^2)$. Notice that we use (\ref{gaugecond0}) to define the gravitational equations of motion in a particular coordinate system and use (\ref{expandh0}) to iteratively solve these in the same coordinates.

Just as (\ref{gaugecond0}) defines the metric decomposition $g_0 + h$ we now define the worldline decomposition $z_0 + \zeta$ through the relation
\begin{equation}
    \big\langle \Delta_1 m ( h, \zeta) a(g_{0}, z_0) + m (g_{0}, z_0) \Delta_1 a (h, \zeta) - \Delta_1 F ( h, \zeta) \big\rangle = 0
    \label{fixaverage0}
\end{equation}
so that the averaged first order perturbed acceleration, effective
mass,  and finite-size effect force vanishes. For pure point particle motion, (\ref{fixaverage0}) reduces
to $\langle \Delta_1 a(h, \zeta) \rangle = 0$.

Writing $\zeta = z_1 + z_2 + O( \epsilon^3)$, averaging
(\ref{EEexample1}) and (\ref{ppexample1}), applying the constraints
(\ref{fixmetric1}) and (\ref{fixmetric2}) to these along with the
definition of the averaging procedure in (\ref{fixaverage0}), and
subtracting from the unaveraged original expressions
(\ref{EEexample1}) and (\ref{ppexample1}) yields the following field
equations for the gravitational variables
\begin{align}
    G(g_{0} ) = {} & T^{epp} ( g_{0}, z_0) - \big\langle \Delta_2 G (  h_1) \big\rangle + 8 \pi G_N \big\langle \Delta_2 T^{epp} (  h_1) + \Delta_2 T^{epp} (z_1) \big\rangle
    \label{bggraveom0} \\
    \Delta_1 G( h_1 ) = {} & 8 \pi G_N \Delta_1 T^{epp} (h_1) + 8 \pi G_N \Delta_1 T^{epp} (z_1)
    \label{gweom1} \\
    \Delta_1 G(h_2)  = {} & - \Delta_2 G( h_1) + \big\langle \Delta_2 G (  h_1) \big\rangle + 8 \pi G_N \Delta_1 T^{epp} (h_2) + 8 \pi G_N \Delta_1 T^{epp} (z_2) + 8 \pi G_N \Delta_1 T^{epp} ( h_1, z_1) \nonumber \\
    & + 8 \pi G_N \left[ \Delta_2 T^{epp} ( h_1)  + \Delta_2 T^{epp} (z_1) - \big\langle \Delta_2 T^{epp} ( h_1) + \Delta_2 T^{epp} (z_1) \big\rangle \right]
    \label{gweom2}
\end{align}
and for the worldline coordinates of the effective point particle
\begin{align}
    m(g_{0}, z_0) a (g_{0}, z_0) = {} & F (g_{0}, z_0) - \big\langle \Delta_2 m (h_1) + \Delta_2 m (z_1)  \big\rangle a (g_{0}, z_0) - m(g_{0}, z_0) \big\langle \Delta_2 a (h_1) + \Delta_2 a (z_1) \big\rangle \nonumber \\
    & - \big\langle \big( \Delta_1 m (h_1) + \Delta_1 m (z_1) \big) \big( \Delta_1 a(h_1) + \Delta_1 a (z_1) \big) \big\rangle + \big\langle \Delta_2 F (h_1) + \Delta_2 F(z_1) \big\rangle
    \label{bgppeom0} \\
        m(g_{0}, z_0) \big( \Delta_1 a (h_1) + \Delta_1 a (z_1) \big) = {} & -  \big( \Delta_1 m (h_1) + \Delta_1 m (z_1)  \big) a (g_{0}, z_0) + \Delta_1 F (h_1) + \Delta_1 F(z_1)
        \label{parteom1} \\
        m(g_{0}, z_0) \Delta_1 a(z_2) = {} & \cdots
        \label{parteom2}
\end{align}
(We have not written the second order equations for the particle coordinates $z_2$ due to their length.)

The equations of motion (\ref{bggraveom0})-(\ref{parteom2}) describe
the inspiral phase of a binary. Since  we did not build into the
expansion any explicit reference to the mass ratio or the relative
velocity these equations may be applicable for describing an
inspiraling binary with arbitrary values for these quantities. Our
hope is that these equations may also be useful for describing the
plunge and possibly merger phases since we can incorporate the
companion-induced tidal effects on the smaller compact object during
the course of its evolution in the strong field region of its larger
mass companion. While such a description using the effective point
particle worldline of \cite{GoldbergerRothstein:PRD73} will surely
break down if the  mass $m$ undergoes tidal disruption or the
horizons of two black holes just merge it may nevertheless be useful
for a description as the merger phase is approached. However, further
investigation is needed to understand these points and the domain of
validity of this system of equations.

The physics contained in these equations is very rich.  The worldline
coordinates, gravitational wave perturbations and background geometry
mutually back-react on each other. This can be seen in
(\ref{bggraveom0}), for example, where the background geometry is
sourced by the leading order stress tensor for the effective
worldline $z_0$, by the averaged second order gravitational waves
(analogous to an averaged Landau-Lifshitz pseudo-tensor) $\sim
\langle \Delta_2 G(h_1) \rangle$, and by the averaged higher order
corrections of the particle stress tensor. As such, the background
geometry cannot and ought not be specified but must be determined
self-consistently with the dynamics of the other gravitational and
particle variables.

Notice also that the equations for $z_0$ are  sourced by averaged
forces involving the first order perturbed coordinates and
gravitational waves. In particular, the worldline coordinates $z_0$
do not follow a geodesic of the background, even when the non-minimal
couplings from (\ref{effppaction0}) are removed, but are determined
by consistently solving (\ref{bgppeom0})-(\ref{parteom2}).

\section{Post-Newtonian expansion for arbitrary masses}
\label{sec:pneftunequal}

The effective field theory paradigm can be applied to any gravitational system where the
typical scales of the gravitational perturbations are much larger
than their sources or scatterers. These conditions can be satisfied
for compact binaries with an arbitrary mass ratio during some portion
of its evolution (e.g., inspiral and ringdown phases). For many
binary systems, such as a neutron star and a $10M_{\odot}$ black
hole, the mass ratio parameter $\eta$ is small suggesting that it can
be used as an additional expansion parameter in the post-Newtonian
approximation. In this section we discuss the role that the mass
ratio
\begin{equation}
    \eta \equiv \frac{ m }{ M } ~~,~~ M \ge m
\end{equation}
has in parameterizing the relevant interactions at the desired order
in the effective field theory approach to the post-Newtonian expansion, PN-EFT.

One purpose for doing this is to provide a better understanding for
the description of binary systems when the mass ratio is tuned to
smaller values. The analytical techniques for describing (nearly)
equal mass binaries are considerably different than those used for
binaries with extreme mass ratios. Whereas the former is based on
assumptions of small
velocities and weak gravitational fields the latter is grounded in
black hole perturbation theory and is applicable to relativistic
binaries with strong gravitational effects. Our approach, described
below, can be regarded as a way to optimize the post-Newtonian
expansion for binary systems with unequal masses by organizing the
perturbative corrections into groups of terms that have numerically
similar magnitudes and relevance at a given post-Newtonian order.

Incorporating $\eta$ into the approximation scheme allows for a more
quantitative estimation of the magnitude of the interactions that
contribute at a given PN order since the
smallness of the mass ratio can suppress some interactions relative
to others. For extreme mass ratio binaries ($\eta \ll 1$) the
interactions accounting for the backreaction on SMBH are
strongly suppressed relative to those describing the leading
order motion of the small compact object, for example.

\subsection{Arbitrary mass ratios in PN-EFT}
\label{sec:nrgr}

Introducing $\eta$ into the post-Newtonian expansion requires
modifying the PN-EFT power counting rules originally developed for
the (nearly) equal mass scenario in \cite{GoldbergerRothstein:PRD73}.
Let $\bz(t)$ and $\bZ(t)$ denote the trajectories of the masses $m$
and $M$, respectively. In terms of these, the binary's center of mass
$\bR$ is described by
\begin{equation}
    \bR (t) = \frac{ \eta}{ 1+\eta} \bz(t) + \frac{ 1}{ 1+\eta} \bZ(t)   \label{CMcoords0}  .
\end{equation}
In the center of mass frame the velocities of the two masses are
related through
\begin{equation}
    \bV(t) = - \eta \bv(t)
\end{equation}
where $\bv = \dot{\bz}$ and $\bV = \dot{\bZ}$ are the 3-velocities of
$m$ and $M$, respectively. From the virial theorem of Newtonian
gravity
\begin{equation}
    m \bv^2 + M \bV^2 = \frac{ 2G_N m M }{  | \bz - \bZ | }
\end{equation}
we find the following relations between the velocities of the masses
and their respective leading order potentials,
\begin{align}
    \bv^2 &= \frac{ G_N M }{ r } \sim \frac{ M }{ m_{pl}^2  r }  \label{virial0}  , \\
    \bV^2 &= \eta \, \frac{ G_N m }{ r } \sim  \eta^2 \, \frac{ M }{ m_{pl}^2  r}   \label{virial1}  .
\end{align}
Here $r$ denotes the magnitude of the orbital separation, $\br = \bz
- \bZ$, and we use the same conventions as
\cite{GoldbergerRothstein:PRD73}. In particular, we use units where
$c=\hbar=1$ and $m_{pl}^{-2} = 32 \pi G_N$.

In PN-EFT there are two kinds of metric perturbations: potential
gravitons $H_{\mu\nu}$ and radiation gravitons $\barh_{\mu\nu}$ \footnote{The use of the word `graviton' should not alarm the reader. While the PN-EFT approach makes use of techniques borrowed from quantum field theory there is nothing quantum in the classical two-body problem. The reader should simply regard a `graviton' as a `gravitational perturbation'.}. The
orbital separation of the masses is much smaller than the
light-crossing time so that each compact object interacts, to leading
order, with the instantaneous potential $H_{\mu\nu}$ generated by the
other mass. At the orbital scale, the emission of gravitational waves
is described by an external, slowly varying, long wavelength
perturbation. Therefore, the full metric is given by
\begin{equation}
    g_{\mu\nu} = \eta_{\mu\nu} + \frac{ H_{\mu\nu} }{ m_{pl} } + \frac{ \barh_{\mu\nu} }{ m_{pl} }
\end{equation}
where the perturbations are canonically normalized by the Planck mass
(equivalently, by $G_N^{-1/2}$).

The power counting rules are given by the virial theorem,
(\ref{virial0}) and (\ref{virial1}), and the ratios of the masses to
the Planck mass, $m / m_{pl} = \sqrt{ \eta v L }$ and $M / m_{pl} =  \sqrt{v L /\eta}$.
Here $L \sim m v r$ is the center of mass angular momentum of $m$ and
the scaling laws of the potential and radiation gravitons are the
same as in \cite{GoldbergerRothstein:PRD73}.

The total action of the system is given by
\begin{equation}
    S_{tot} [ \bz, \bZ, H, \barh] = S [ \eta + H/m_{pl} + \barh/m_{pl} ] + S_{epp} [ \bz, H, \barh] + S_{epp} [ \bZ, H, \barh]
\end{equation}
where the first term is the gauge-fixed Einstein-Hilbert action and
the last two describe the masses (treated as effective point particles \cite{GoldbergerRothstein:PRD73}) coupled
to potential and radiation gravitons. There are many non-minimal
couplings in the effective point particle Lagrangians that describe
the effects from companion induced tidal moments but these are higher
order effects that we will ignore for most of our discussion.

The larger mass $M$ couples locally to the potential graviton via
factors of
\begin{equation}
    H_{\mu\nu} \big( t, \bZ(t) \big) = H_{\mu\nu} \big( t, -\eta \bz(t) \big)
\end{equation}
in the point particle Lagrangian. The mass ratio parameter appears in
the argument of $H_{\mu\nu}$ and implies that
all $n$-point functions of potential gravitons will give rise to
contributions proportional to products of
\begin{equation}
    \big\langle H_{\mu\nu} (t, \bz) H_{\alpha \beta} (t, \bZ) \big\rangle \propto \frac{ 1}{ |\bz - \bZ| } = \frac{1}{ (1 + \eta ) | \bz | } ,
\end{equation}
which do not scale homogeneously with $\eta$. However, by expanding
the potential graviton in multipoles about the center of mass, which
we take to be at the origin,
\begin{equation}
    H_{\mu\nu} (t, \bZ(t)) = H_{\mu\nu} (t, \bzero) + Z^i(t) \partial_i H_{\mu\nu} (t, \bzero) + \frac{1}{2} Z^i(t) Z^j(t) \partial_i \partial_j H_{\mu\nu} (t, \bzero) +  \cdots
\end{equation}
we see that each term above now scales as a definite power of the
mass ratio since $(Z^i \partial_i)^p H_{\mu\nu} \sim \eta^p
H_{\mu\nu}$ for some non-negative integer $p$. As in
\cite{GoldbergerRothstein:PRD73} the radiation gravitons are also
expanded in multipoles
\begin{align}
    \barh_{\mu\nu} \big( t, \bz(t) \big) &= \barh_{\mu\nu} (t, \bzero) + z^i \partial_i \barh_{\mu\nu} (t, \bzero) + \frac{1}{2} z^i z^j \partial_i \partial_j \barh_{\mu\nu} (t, \bzero) + \cdots , \\
    \barh_{\mu\nu} \big( t, \bZ(t) \big) &= \barh_{\mu\nu} (t, \bzero) + Z^i \partial_i \barh_{\mu\nu} (t, \bzero) + \frac{1}{2} Z^i Z^j \partial_i \partial_j \barh_{\mu\nu} (t, \bzero) + \cdots   .
\end{align}
The power counting rules imply that $z^i \partial_i
\barh \sim v \barh$ whereas $Z^i \partial_i \barh \sim \eta v \barh$.

As an example of the new power counting rules, it is straightforward to show that companion-induced tidal effects (i.e. the effacement of internal structure) for the smaller mass $m$ first appear at order $\eta^4 v^{10}$ and for the larger mass $M$ at order $\eta^1 v^{10}$. In the EMRI limit, where $v \sim 1$, these agree with earlier results \cite{Galley:EFT1, Zerilli:PRD2}.

\subsection{Estimates of post-Newtonian orders for the number of gravitational wave cycles}


Current and future generation gravitational wave interferometers will
be capable of extraordinarily precise measurements partly because the
interesting signals, which are buried in a sea of detector noise, can
be extracted using a technique called matched filtering \cite{Finn:PRD46, FinnChernoff:PRD47}. Matched filtering is most efficient when the sought after signal is
known or modeled with a sufficient accuracy. For gravitational wave
interferometers the signal is the waveform
\begin{equation}
    h(t) = F_+ h_+ (t) + F_ \times h_\times (t)
\end{equation}
given in terms of the $+$ and $\times$ polarizations of the
gravitational wave and the pattern functions $F_{+, \times}$, which
depend on the geometry of the interferometer and the direction to the signal source. For inspiraling
sources, these detectors will accumulate many oscillations of the
gravitational wave so that the phase of $h(t)$ must be known to a
high degree of precision.

In this section we will provide rough estimates for determining the
post-Newtonian orders, as a function of the (unequal) masses of the binary, necessary to construct templates that will be
sufficient for precisely estimating the parameters of a detected
source of gravitational waves.

We begin by noting that the
change in the number of cycles spent in a logarithmic interval of
frequency can be computed within the post-Newtonian approximation and
has the generic form \cite{Cutler:PRL70}
\begin{equation}
    \frac{ d N_{cyc} }{ d \ln f } = \frac{ 5}{ 96 \pi \xi^{5/3} } \frac{ f^{-5/3} }{ \eta}  \sum_{p, q = 0 } ^\infty c_{pq} \eta^p v^q (f)
    \label{dcyclesdf0}
\end{equation}
where $f$ is the frequency of the emitted gravitational radiation,
$c_{pq}$ are coefficients that we  assume to be $O(1)$ \footnote{We
realize that this is not true for every $p$ and $q$ (e.g., see the PN expansion of the gravitational wave phase in \cite{Blanchet:LRR}). However,
we are unaware of any natural scaling arguments that would otherwise indicate
the magnitudes of these coefficients.}, and $\xi = G (m+M)
\pi$ in units where $c=1$. The total number of cycles spent in a
given frequency interval is found by integrating both sides of (\ref{dcyclesdf0}) so that
\begin{align}
    N_{cyc} (f) & =  \frac{5  }{96 \pi}   \sum_{p=0}^\infty c_{p5} \eta^{p-1} \ln \left( \frac{ f_{isco} }{ f } \right) + \frac{15}{96\pi} {\sum_{p,q=0}^\infty}^\prime \frac{ c_{pq} }{ q-5 } \xi^{(q-5)/3} \eta^{p-1} \left( f_{isco}^{(q-5)/3 } - f^{(q-5)/3} \right)
    \label{cycles0} \\
    & \equiv \sum_{p,q=0}^\infty N_{cyc}^{(p,q)} (f)
\end{align}
where we have used Kepler's Law $v^3 = \xi f$ to relate the
(circular) relative velocity to the gravitational wave frequency and
the prime on the second summation means that the $q=5$ term is
omitted. We take the upper limit of the frequency interval to be the
gravitational wave frequency at the innermost stable circular orbit
(ISCO). The ISCO demarcates the boundary between the quasi-circular
inspiral and plunge phases of the binary's evolution.

It is widely believed that (\ref{dcyclesdf0}) should be computed to
3.5PN order for the following reason \cite{Maggiore}. Using Kepler's
Law, the leading order term ($p,q=0$) scales as $v^{-5}$ so that the
term in the sum of (\ref{dcyclesdf0}) that goes as $v^5$ (2.5PN) will
give an $O(1)$ correction to $N_{cyc}$ and, equivalently, the phase
of $h(t)$. However, an $O(1)$ correction implies that the template
has ``slipped" from the signal by an $O(1)$ number of cycles thereby
reducing the chances of {detecting the signal using matched
filtering}. Instead, one should compute $N_{cyc}$ to 3PN, or better
yet 3.5PN for higher precision. Furthermore, these higher orders
vanish in the limit that $v\to0$ implying that these corrections are
potentially small.

\begin{figure}[t]
    \centering
    \subfigure[~$\delta = 0.1$]{
        \includegraphics[scale=.25]{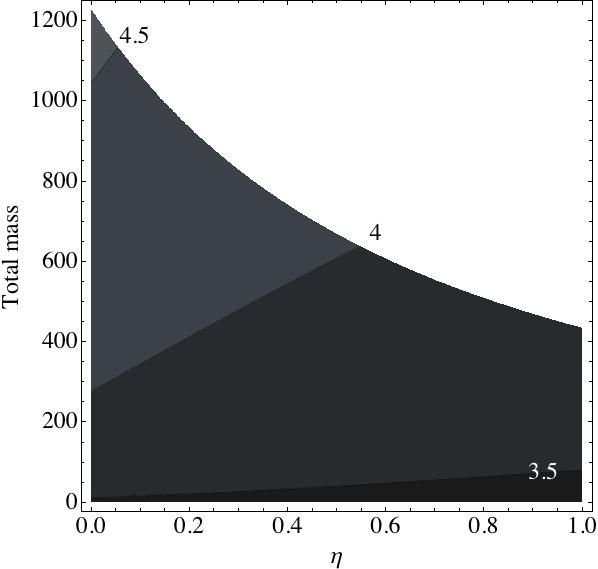}
        \label{fig:subfig1} }
    \subfigure[~$\delta = 0.05$]{
        \includegraphics[scale=.25]{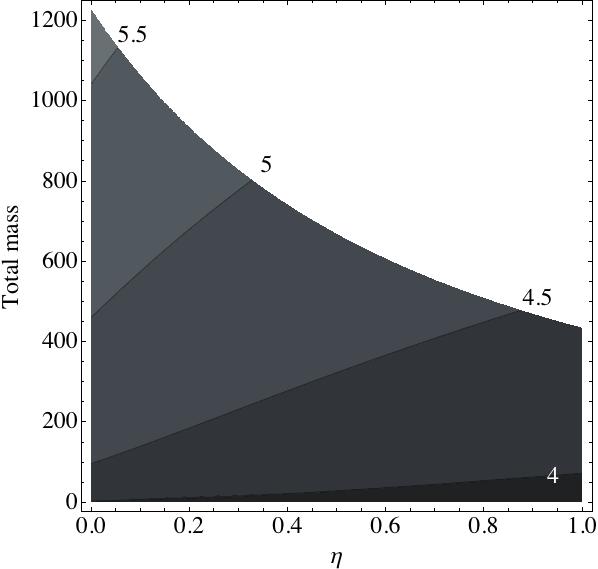}
        \label{fig:subfig2} }
    \subfigure[~$\delta = 0.01$]{
    \includegraphics[scale=.25]{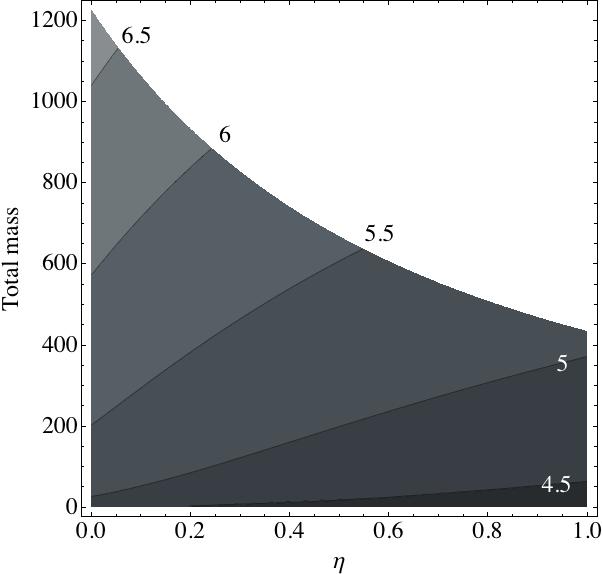}
    \label{fig:subfig3} }
    \caption[Optional caption for list of
figures]{Contour plots showing the post-Newtonian orders $q/2$ needed
to estimate $R_{cyc}^{(q)}$ with a precision (relative to the 2.5PN contribution) of \subref{fig:subfig1}
$10\%$, \subref{fig:subfig2} $5\%$, and \subref{fig:subfig3} $1\%$ as
a function of the total mass (in $M_\odot$) and the mass ratio $\eta$. The white
areas indicate those binaries with masses that have ISCO frequencies
lower than $10$Hz, which we regard here as not detectable by LIGO.}
    \label{fig:percentages}
\end{figure}

The $p,q=0$ term in (\ref{cycles0}) denotes the leading order (i.e.,
Newtonian) prediction for $N_{cyc}$ and is often used for making
rough estimates of the number of cycles spent in LIGO's bandwidth.
The remaining terms are the post-Newtonian corrections from higher
order gravitational, relativistic, spin, radiation reaction, and
finite size effects. If we regard LIGO's lowest detectable frequency
to be approximately 10 Hz \footnote{The low frequency part of LIGO's bandwidth is not
sharply defined at 10 Hz but the noise curve does rise rather steeply for
frequencies near 10 Hz.} then the ratio
\begin{equation}
    R_{cyc}^{(q)} (10 {\rm \, Hz}) \equiv \frac{ N_{cyc}^{(p, q)} (10 {\rm \, Hz}) }{ N_{cyc}^{(p,5)} (10 {\rm \, Hz})} \approx \frac{ 3 \xi^{(q-5)/3} }{ q-5 } \frac{ f_{isco}^{(q-5)/3} - 10 ^{(q-5)/3} }{ \ln ( f_{isco}/10 ) }  ,
    \label{ratio0}
\end{equation}
with $q>5$ and $c_{pq} = O(1)$, gives an approximate measure of the corrections to the $O(1)$ number of cycles (i.e., the 2.5PN correction) from higher PN terms. Note that the ratio is independent of $p$.

From (\ref{ratio0}) we can estimate the approximate PN order, with a
specified precision, that (\ref{cycles0}) should be expanded through.
If $\delta$ denotes this precision then the PN order is approximately
given by the solution of
\begin{equation}
    R_{cyc}^{(q)}(10 {\rm \, Hz}) = \delta
    \label{ratio1}
\end{equation}
for $q/2$. A more accurate treatment should be given by more
quantitative means \cite{Thorne:RevModPhys52, Cutler:PRL70} but since $N_{cyc}$ is
currently known only through 3.5PN \cite{Blanchet:LRR, Maggiore} then we will continue with this
somewhat crude estimation.

Figures \ref{fig:subfig1}-\ref{fig:subfig3} show contour plots of the
solutions to (\ref{ratio1}) as a function of the binary's total mass
and the mass ratio for $\delta = 0.1, 0.05$, and $0.01$,
respectively. For a precision of $\approx 10\%$ Figure
\ref{fig:subfig1} indicates that 3.5PN expressions are adequate for
equal mass binaries with a total mass as high as $\sim 80 M_{\odot}$
but that 4PN expressions may be more suitable for total masses
between $\sim 80 M_\odot - 430 M_\odot$. Equal mass binaries with a
total mass $\gsim 430 M_\odot$ emit gravitational waves at the ISCO
with frequencies less than 10Hz and therefore do not enter LIGO's
bandwidth. At a precision of $1\%$ Figure \ref{fig:subfig3} shows
that for a binary with a mass ratio of $1/10$ and a total mass of
$1000 M_\odot$ that one may need PN expressions to order 6.5PN. {For
a precision of $10\%$ Figure \ref{fig:subfig1} implies that the same
system may require 4.5PN corrections. Such high PN orders for this
particular binary should be expected since its waveform is in LIGO's
observable bandwidth for a short frequency interval.}

The plots in Figure (\ref{fig:percentages}) should be regarded as
providing crude estimates for determining the PN orders needed for
LIGO to measure the number of cycles in its bandwidth with a given
precision. Of course, these estimates can be justified only after the
PN expansion of $N_{cyc}$ has been performed to higher orders in conjunction with a more comprehensive analysis.
Nevertheless, we believe that Figure (\ref{fig:percentages}) and the considerations of this section can
serve as a motivation for calculating higher order corrections in the PN
expansion for binaries with unequal masses.

\section{Further developments}

\subsection{Gravitational perturbation theory with self-consistent backreaction}

{The GP-SCB approach provides a general framework to build
perturbation theories that are specific to the assumptions and input
being considered. We demonstrated in Section \ref{sec:specialcases}
how the well-known post-Newtonian expansion and perturbation theory
for EMRIs can be derived in the GP-SCB formalism.}

In Section \ref{sec:adiabaticbkrxn} we used the GP-SCB approach to
introduce an adiabatic expansion with backreaction for arbitrary
masses and velocities.  Perhaps the most pertinent issue is
determining the domain of validity of the equations of motion
(\ref{bggraveom0})-(\ref{parteom2}). {In particular, the degree of
accuracy of those equations needs to be determined by the data
analysis requirements set by gravitational wave interferometers.}
Can these system of equations adequately capture the plunge and
possibly merger phases? How does the mass ratio and relative velocity
affect the answer to these questions, if at all? We hope to address
these questions in some more detail in future papers
\cite{GalleyHu:SCB1, GalleyHu:SCB2}.

A related issue is obtaining the solutions to such complicated
backreaction equations. Since all of the degrees of freedom in the
example of Section \ref{sec:adiabaticbkrxn} are interacting
non-trivially and non-linearly with other variables it would seem
that there is a rich variety of processes that makes calculating the
number of cycles that a binary stays in a detector's bandwidth, for
example, difficult to perform and subsequently to compare with known
results, both analytical and numerical. Furthermore, in many cases
there are isometries associated with the background geometry that can
be exploited to assist in developing (semi-)analytical solutions.
This is the case with the post-Newtonian method applied to the
Minkowski spacetime, which is maximally symmetric, and with
EMRI-perturbation theories off a Kerr background, which is
axisymmetric. However, there does not seem to be any isometry
associated with the background metric $g_{0\mu\nu}$ since its
evolution is determined in part by sources lacking any specific
symmetry; see (\ref{bggraveom0}). Consequently, we will likely be
forced to introduce some additional assumptions, but hopefully not as
drastic as weak field, slow motion or small mass ratio, that could
make it easier to find solutions beyond these simpler and more
accessible regimes and provide new insights into processes hitherto
too difficult to comprehend analytically.

There have been several approaches that are based on an adiabatic
approximation from a two-time scale separation for EMRI binaries
\cite{HindererFlanagan:PRD78, Mino:ProgTheorPhys113,
Mino:ProgTheorPhys115, Mino:CQG22_1, Mino:CQG22_2,
PoundPoisson:PRD77}. While our method is not restricted to extreme
mass ratios, it is nevertheless more involved because the background
geometry evolves with the backreaction from both the stress energy of
the compact object (with finite size effects due to the extended
nature of the mass) and the effective stress energy of the
gravitational waves.

The generic framework for building  perturbative expansions,
originally developed in \cite{Anderson:PRD55}, may be useful for
generating other approximations in addition to the one discussed in
Section \ref{sec:adiabaticbkrxn}. Depending on the dynamical regime
of the two-body dynamics it may be possible to develop more
perturbation theories suitable for other scenarios involving
gravitational binaries including the head-on collision of two black
holes, the merger of comparable mass compact objects, etc. One
obvious perturbation theory that can be constructed would use the
effective point particle description for both masses of the binary
and derive the equations of motion in a similar manner as in Section
\ref{sec:adiabaticbkrxn}. Such a theory might lend itself useful to
more practical schemes for solving the backreaction equations. We
plan to investigate these and other issues associated with the
self-consistent approach in future papers \cite{GalleyHu:SCB1, GalleyHu:SCB2}.

\subsection{Post-Newtonian effective field theory for arbitrary mass ratios}

In Section \ref{sec:pneftunequal}  we provided crude estimates
indicating the PN order that the number of gravitational wave
cycles falling into LIGO's bandwidth should be calculated through for
several given precisions. The PN order depends non-trivially on the
binary's mass ratio and total mass. For a binary composed of $10
M_\odot$ and $100 M_\odot$ compact objects, Figure \ref{fig:subfig2}
suggests that one may need to calculate $N_{cyc}$ to 4.5PN to achieve
a $\sim 5\%$ precision relative to the 2.5PN correction. We also
find that for a constant total mass the PN order increases as the
mass ratio decreases. Meanwhile, for a constant mass ratio the PN
order increases as the total mass increases. Both results are
expected on intuitive grounds.

Justifying these estimates requires computing  to higher PN orders
beyond the current 3.5PN expressions. In the PN-EFT approach, such
calculations can be accomplished systematically and relatively
efficiently compared to standard formalisms. Indeed, where higher PN
orders are needed new physical interactions can manifest, including
finite size effects from companion-induced tidal deformations (i.e.,
the non-effacement of internal structure), spin effects with
radiation reaction, etc. We intend to investigate some of these
features in future work.

\acknowledgments

This work is supported in part by NSF grants PHY-0801368
and PHY-0801213
to the University of Maryland. The gravitational
perturbation theory with self-consistent backreaction described here
is based on a part of CRG's PhD thesis work at the University of Maryland,
supported in part by NSF grant PHY-0601550. CRG is indebted to Paul
Anderson for clarifying many important details regarding his
gauge-invariant description of effective stress-energy tensors for
gravitational waves and for providing us comments on a previous version of this paper. BLH wishes to thank Professors Larry Horowitz,
Martin Land and Ioannis Antoniou and other organizers of the 2008
International Conference on Classical and Quantum Relativistic
Dynamics of Particles and Fields for their hospitality.


\bibliographystyle{physrev}
\bibliography{imri_bib}

\setlength{\parskip}{1em}



\end{document}